\documentclass[10pt,twocolumn,letterpaper]{article}

\usepackage{3dv}

\usepackage{times}
\usepackage{epsfig}
\usepackage{amsmath}
\usepackage{amssymb}
\usepackage{amsfonts}
\usepackage{color}
\usepackage{url}
\usepackage{breakurl}
\usepackage{graphicx}
\usepackage{subfigure}
\graphicspath{{Images/}}
\usepackage{float} 

\newcommand{\todo}[1]{}
\newcommand{\ivo}[1]{}

\newcommand{\na}{N\!A} 

\newcommand{\ri}{n}    
\newcommand{\fl}{f}    
\newcommand{\fn}{F}    
\newcommand{\magn}{M}  
\newcommand{\diam}{D}  

\newcommand{\mum}{\mu m}
\newcommand{\mm}{mm}

\usepackage[pagebackref=true,breaklinks=true,letterpaper=true,colorlinks,bookmarks=false]{hyperref}

\threedvfinalcopy 

\def\threedvPaperID{86} 

\ifthreedvfinal\fi
\begin{document}

\title{Light-Field Microscopy with a Consumer Light-Field Camera}

\author{Lois Mignard-Debise\\
INRIA, LP2N\\
Bordeaux, France\\
{\tt\small \url{http://manao.inria.fr/perso/~lmignard/}}
\and
Ivo Ihrke\\
INRIA, LP2N\\
Bordeaux, France\\
}

\maketitle


\begin{figure*}[t]
	\includegraphics[width=\textwidth]{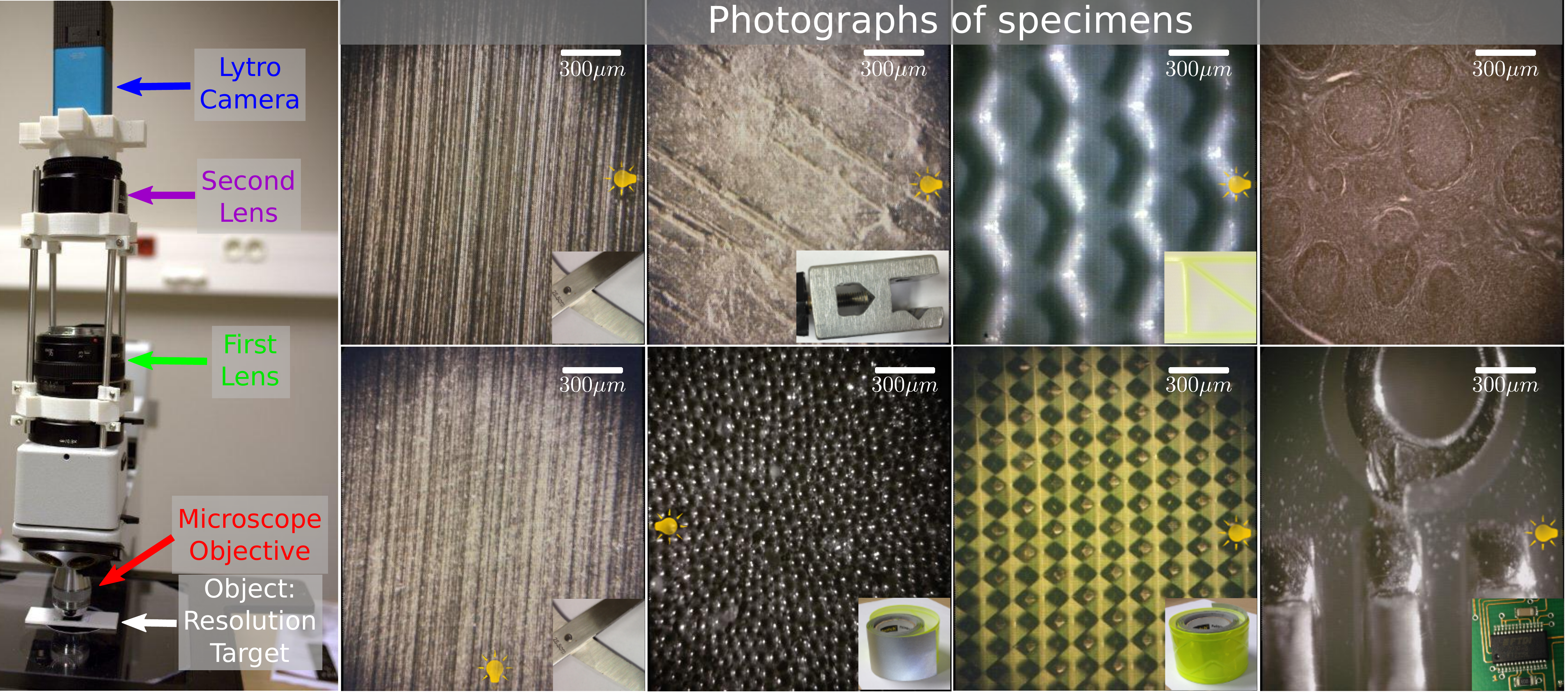}
        %
        \caption{{\em Left:} Microscope setup with the Lytro camera on
          top of two additional SLR lenses. Images of several samples
          have been taken under different illumination
          conditions. Magnification is 3.0. {\em Top and bottom left:}
          brushed steel from scissors blade with lighting from the
          right and bottom. {\em Top middle left:} Scratched surface of a
          piece of metal. {\em Bottom middle left:} Plastic surface with
          highly retro-reflective properties. The material is made of
          micro bubbles of transparent plastic that are invisible to
          the naked eye. {\em Top middle right:} Fabric with a hexagonal
          structure. The lighting is coming from the side and casts
          shadows and strong highlights on the three-dimensional
          structure of the fabric. {\em Bottom middle right:} Highly
          retro-reflective material from security reflective tape.
          {\em Top right:} Tonsil tissue with bright field illumination.
          {\em Bottom right:} Pins of a electronic component on a circuit board.}
          \vspace{-3mm}
        \label{fig:samples_results}
\end{figure*}


\begin{abstract}

   We explore the use of inexpensive consumer light-field camera
   technology for the purpose of light-field microscopy.
   Our experiments are based on the Lytro (first generation)
   camera. Unfortunately, the optical systems of the Lytro and those
   of microscopes are not compatible, leading to a loss of light-field
   information due to angular and spatial vignetting when directly
   recording microscopic pictures.  We therefore consider an
   adaptation of the Lytro optical system.

   We demonstrate that using the Lytro directly as an ocular
   replacement, leads to unacceptable spatial vignetting. However, we
   also found a setting that allows the use of the Lytro camera in a
   virtual imaging mode which prevents the information loss to a large extent.
   We analyze the new virtual imaging mode and use it in two different
   setups for implementing light-field microscopy using a Lytro
   camera.  As a practical result, we show that the camera can be used
   for low magnification work, as e.g. common in quality control,
   surface characterization, etc. We achieve a maximum spatial
   resolution of about $6.25\mu m$, albeit at a limited SNR for the side
   views.

\end{abstract}

\section{Introduction}
\label{sec:intro}

Light-field imaging is a new tool in the field of digital
photography. The increasing interest is shown by the recent
development of several hardware systems on the consumer market (Lytro,
Raytrix, Picam), and applications in the research domain
(stereo-vision, panoramic imaging, refocusing).
The commercial systems are reliable, functional and
inexpensive. However, they are designed for the imaging of macroscopic
objects. In this article, we explore the use of commercial light-field
cameras for microscopic imaging applications.

The light-field microscope has been introduced and improved by Levoy
et al.~\cite{levoy2006light,Levoy:09,Broxton:13}. While its conceptual
details are well understood, its practical implementation relies on
the fabrication of a custom micro-lens array, which presents a hurdle
for experimenting with the technology. In this article, we demonstrate
the use of the Lytro camera, an inexpensive consumer-grade light-field
sensor, for microscopic work. We achieve an inexpensive and accessible
means of exploring light-field microscopy with good quality, albeit at
a reduced optical magnification.

As we show, the major problem in combining the Lytro and additional
magnification optics (in addition to f-number matching), is the loss
of information due to spatial vignetting.  Our main finding is the
possibility of using the Lytro in what we term an {\em inverse
regime}: in this setting the camera picks up a virtual object that is
located far behind its imaging optics. To our knowledge, this is the
first time that such a light-field imaging mode is described.

We investigate two different setups based on this inverse regime that
do not suffer from spatial vignetting: 1) Our first option enables the
use of the Lytro camera in conjunction with an unmodified microscope
by designing an optical matching system. 2) The second option uses the
Lytro behind a standard SLR lens in a macrography configuration to
achieve macro light-field photography.

The paper is organized as follows. In Section~\ref{sec:background}, we
study the compatibility of both optical systems involved in the
imaging process: the light-field camera and the microscope. We discuss
the implications of the combination of both systems in terms of both
spatial and angular resolution. We then explore the Lytro main optical
system and show how to adapt it to the microscopic imaging context,
Sects.~\ref{sec:impladdoptics} and~\ref{sec:implrmoptics}. Finally, we
evaluate and compare the different solutions and present application
scenarios, Sect.~\ref{sec:results}.

\section{Related Work}
\label{sec:related}
\textbf{Light-field imaging}
%
%
requires the acquisition of a
large number of viewpoints of a single scene. Two types of approaches
exist, either using multiple sensors or a single sensor in conjunction
with temporal or spatial multiplexing schemes.

Taking a picture with a conventional camera is similar to a 2D slicing
of the 4D light-field. Repeating this operation with a planar array of
cameras offers sufficient data to estimate the light-field
function. Calibration, synchronization, as well as the available
bandwidth of the camera hardware are the major determining features of
this approach. More details can be found, e.g.,
in~\cite{yang2002real,wilburn2005high}.  The strongest limitation of
this approach for microscopic applications is the large size of the
corresponding setups.

An alternative method for light-field capture is to multiplex the
different views onto a single sensor.

{\em Temporal multiplexing} is based on taking several pictures over time
after moving the camera around static scenes. The movements can be a
translation or rotation of the camera. Alternatively,
mirrors \cite{ihrke2008fast,taguchi2010axial} can be moved to generate
additional virtual viewpoints. Another alternative implementation are
dynamic apertures~\cite{liang2008programmable}.

{\em Spatial multiplexing} allows to record dynamic
scenes. Parallax barriers and integral
imaging~\cite{lippman1908epreuves} are historically the first
approaches to spatially multiplex the acquisition of a light-field,
trading spatial resolution for angular resolution. A modern
elaboration of this approach where the sensor and a micro-lens array
are combined to form an in-camera light-field imaging system is the
Hand-Held Plenoptic Camera~\cite{ng2005light}. Alternatively, sensor
masks~\cite{Veeraraghavan:07,Veeraraghavan:08} or a light pipe~\cite{Manakov:13} can be
arranged such that in-camera light-fields can be
recorded. Other methods use external arrays of
mirrors instead of lenses~\cite{lanman2006spherical}, or external lens
arrays~\cite{georgiev2006light}.

\textbf{Microscopy}
is a vast subject and many different illumination and observation
schemes have been developed in the past. A general overview is given
in~\cite{Murphy:12}; a comprehensive review of microscopy techniques,
including light-field microscopy, for the neuro-sciences can be found
in~\cite{Wilt:09}. Light-field microscopy was introduced by Levoy et
al.~\cite{levoy2006light} and later augmented with light-field
illumination~\cite{Levoy:09}. Recently, addressing the large spatial
resolution loss implicit in LFM, the group has shown that
computational super-resolution can be achieved outside the focal plane
of the microscope~\cite{Broxton:13,cohen2014enhancing}. Another super-resolution scheme
is combining a Shack-Hartmann wavefront sensor and a standard 2D image
to compute a high-resolution microscopic light-field~\cite{Lu:13}.
LFM has been applied to polarization studies of mineral
samples~\cite{Oldenbourg:08} and initial studies for extracting depth
maps from the light-field data have been performed in microscopic
contexts~\cite{Lee:12,Thomas:13}. Most of the work today uses the same
optical configuration that was introduced in the original
implementation~\cite{levoy2006light}.
With this article, we aim at providing an inexpensive means of
experimenting with LFM.


\section{Background \& Problem Statement}
\label{sec:background}

\subsection{Light-field Microscopy}
\label{sec:LFM}
\begin{figure}[h]
 \begin{center}
   \includegraphics[width=\linewidth]{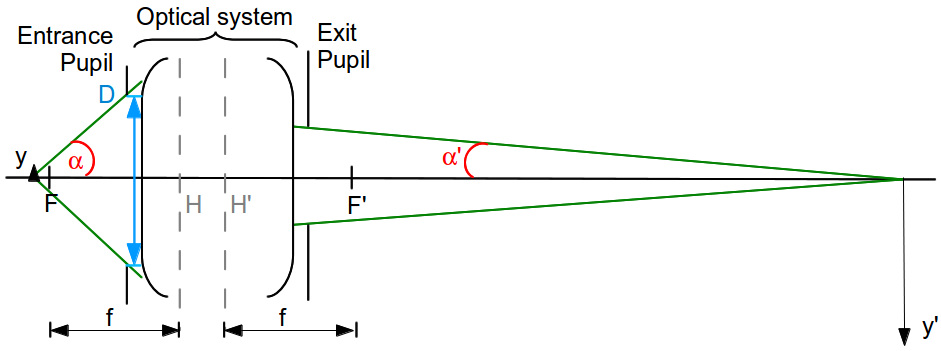}
 \end{center}
 \vspace{-3mm}
 \caption{Properties of an optical
   system. An optical system is defined by its focal length $\fl$,
   its principal planes $H$ and $H'$, and the diameter
   $\diam$ of its entrance pupil. Two conjugate planes define a unique
   magnification $\magn$, the ratio of the image size $y'$ over the
   object size $y$. All optical equations can be found in
   \cite{born1999principles}.}
 \label{fig:optical_definition} 
\end{figure}

The main function of an optical microscope is to magnify small objects
so that they can be observed with the naked eye or a camera
sensor. Light-field capabilities such as changing the viewpoint,
focusing after taking the picture, and achieving 3D reconstruction of
microscopic samples rely on the number of view points that can be
measured from a scene. This number is directly linked to the object-side
numerical aperture $\na_o$ of the imaging system that is used as an
image-forming system in front of the micro-lens array of the
light-field sensor and the micro-lens f-number. The object-side
numerical aperture is defined as
\vspace{-2mm}
\begin{equation}
\na_o = \ri\ sin (\alpha),
\vspace{-2mm}
\end{equation}
where $\ri$ is the index of the material in object space (usually air,
i.e.  $\ri=1$).  The numerical aperture quantifies the extent of the
cone of rays originating at an object point and being permitted into the
optical system (see Fig.~\ref{fig:optical_definition}). Microscope
objectives usually have a high $\na_o$, because it is directly linked
to better optical resolution and a shallower depth of field. A high
$\na_o$ is also important for light-field microscopy as the base-line
of the light-field views is directly linked to it.

Details on how to design a light-field microscope can be found
in~\cite{levoy2006light}. The most important aspect is that the
f-number of the micro-lens array matches the f-number of the
microscope objective. The f-number $\fn$ of any optical
system is defined as the ratio of its focal length $\fl$ over the diameter D of
its entrance pupil (see Fig.\ref{fig:optical_definition})
\vspace{-2mm}
\begin{equation}
\fn = \dfrac{ \fl }{ \diam }.
\vspace{-2mm}
\label{equ:fnumber}
\end{equation}
For a microscope of magnification $\magn_{microscope}$ and numerical
aperture $\na_o$, a more appropriate equation taking the finite image distance into account can be
derived~\cite{born1999principles} from Eq. \ref{equ:fnumber}:
\vspace{-2mm}
\begin{equation}
\fn_{microscope}= \dfrac{ \magn_{microscope} }{ 2\na_o }.
\vspace{-2mm}
\label{equ:microscope_fnumber}
\end{equation}
The majority of microscope objectives has an f-number between 15 and
40.  In our experiments, we use a $10\times$ objective
with an f-number of $\fn=20$.

\subsection{Lytro Features}
\begin{figure}[h]
 \begin{center}
   \includegraphics[width=0.49\linewidth]{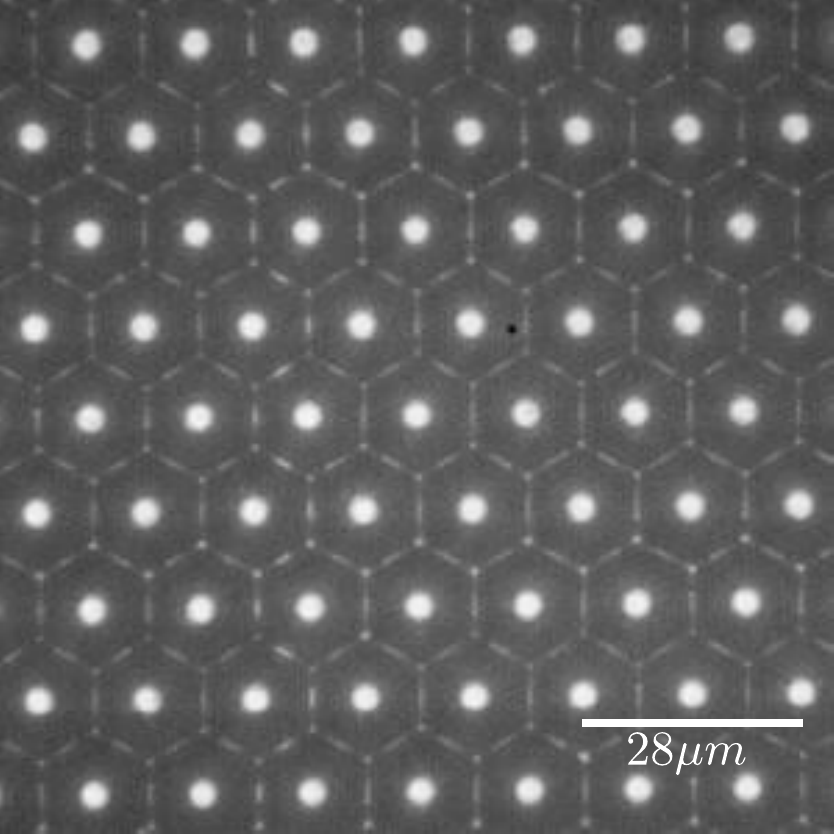}
   \includegraphics[width=0.49\linewidth]{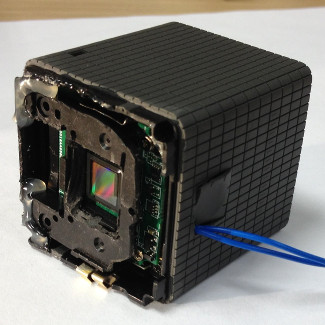}

 \end{center}
 \vspace{-3mm}
 \caption{{\em Left:} Image of the micro-lens array taken
   with a microscope with top illumination. We can see the
   hexagonal structure of the micro-lens array. The bright dot in the
   middle of each hexagon is an image of the light source reflected by
   the surface of the micro-lens. {\em Right:} The Lytro camera without its optics.}
 \label{fig:mla_closeup2} 
\end{figure}

The Lytro camera is made of an optical system that is forming an image
in the plane of a micro-lens array that is, in turn, redirecting the
light rays to a sensor. It has a $3280\times3280$ pixel CMOS sensor
with 12-bit A/D and $1.4 \mum \times 1.4 \mum$
pixels~\cite{miloushnet}. Each micro-lens has a diameter of $14 \mum$
which is equivalent to 10 pixels. The micro-lenses are packed on a
hexagonal lattice (see Fig.~\ref{fig:mla_closeup2}~(left)). The
effective spatial resolution is therefore 328$\times$328 pixels
whereas the angular sampling rate is $10\times10$ values.  The Lytro's
main objective lens has a fixed f-number of $\fn=2$ and features an
$8\times$ optical zoom. We explore its potential as an imaging
parameter in Sect.~\ref{sec:impladdoptics}.  Another important feature
of the objective lens is that it can focus from $0\mm$ to infinity.
\subsection{F-number Mismatch}
\label{sec:mismatch}
A prerequisite for non-vignetted imaging (see also
Sect.~\ref{sec:impladdoptics}) is that the f-number of the micro-lens
array and that of the microscope objective match. This is the solution
employed in conventional light-field
microscopy~\cite{levoy2006light,Oldenbourg:08,Levoy:09}.  A custom
$\fn=20$ micro-lens array (MLA) is typically employed as it is compatible with a
large number of existing microscope objectives. However, this $\fn=20$ MLA
is not readily available and has to be custom manufactured.

Since the Lytro is designed for macroscopic imaging, its f-number
($\fn=2$) is not adapted to the microscopic situation. If we were to
use the Lytro micro-lens array as is, the angular sampling in one
direction would be divided by the ratio of the f-number of the two
systems and only one pixel would be lit under each micro-lens (instead
of approximately one hundred) (see Fig.~\ref{fig:vignetting}~(bottom)). This
would remove any interest for light-field purposes since only a single
view would be recorded and $99\%$ of the sensor would remain unused,
see Sect.~\ref{sec:impladdoptics} for examples.

Therefore, in order to successfully use the Lytro camera for
microscopic imaging, the f-numbers of the two optical systems need to
be adapted. There is, however, a trade-off. The optical invariant,
a fundamental law of optics (see e.g. \cite{born1999principles}),
states that for two conjugate planes, the product between the sine of the angle at which
light rays reach a conjugate plane ($\alpha$ and $\alpha'$) and the
size of the object in this plane ($y$ and $y'$) is equal at both
planes.
\vspace{-2mm}
\begin{equation}
  y\ \ri \ sin(\alpha)\ =\ y'\ \ri'\ sin(\alpha'),
  \vspace{-2mm}
  \label{equ:Abbe}
\end{equation}
 $n$ and $n'$ are the optical index of the media on both sides of the optical system. For air, the index is equal to 1.

We therefore opt for an optical demagnification scheme, decreasing
$y'$, to increase the angular size of the cone of light rays $\alpha'$
that is incident on the Lytro's light-field sensor.
%
%
Theoretically, we need to divide the microscope objective's f-number
by 10 to reach the same f-number as the Lytro camera. An immediate
consequence from Eq.~\ref{equ:microscope_fnumber} is that the
combination of all optical elements must therefore have a
magnification divided by 10, i.e. we are aiming to convert the system
to unit-magnification. Due to the small size of the Lytro's
micro-lenses, the optical resolution of the system is still satisfactory,
even at this low magnification (see Sect.~\ref{sec:results}).

The magnification of the combined system $\magn_{final}$ can be
written as the product of the magnifications of each individual
system:
\vspace{-1mm}
\begin{equation}
\magn_{final} = \magn_{microscope} \magn_{lens_1} ...\  \magn_{lens_N} \magn_{Lytro},
\vspace{-1mm}
\label{equ:magnification}
\end{equation}

where $\magn_{lens_i,\,\,i=1..N}$ indicates the magnification of $N$
to-be-designed intermediate lens systems. The microscope objective has
a fixed magnification of $\magn_{microscope} = 10$,
whereas the lowest magnification setting of the Lytro has a value of
$\magn_{Lytro}=0.5$. The resulting $\magn_{final} = 5$ without
additional optical components ($N=0$) is too large to prevent angular
information loss.

\begin{figure}[h]
 \begin{center}
   \includegraphics[width=\linewidth]{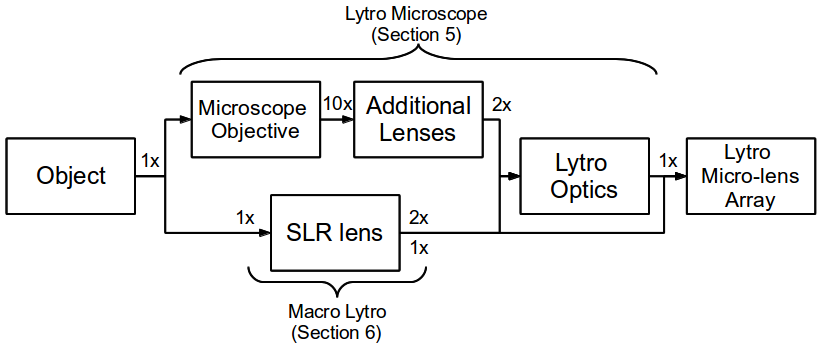}
 \end{center}
 \vspace{-3mm}
 \caption{The diagram shows our two proposed solutions to achieve low magnification. The first option (first row) keeps the camera and the microscope intact. The second option (second row) replaces the microscope with an SLR lens.}
 \label{fig:boxdiagram} 
\end{figure}

We explore two different options (see Fig.~\ref{fig:boxdiagram}) to
implement the adapted system.  The first option (see
Sect.~\ref{sec:impladdoptics}) is to demagnify the image of the
microscope with additional lenses ($N = 2$). This solution lets us use
the microscope and the Lytro camera unmodified.
The second option is to remove the microscope, replacing it by an SLR
lens in macro-imaging mode. Here, we compare a setup with and
without the Lytro optics (see Sect.~\ref{sec:slr_lens}).

\section{Sensor Coverage}
\label{sec:vignetting}
\paragraph{Vignetting:}
When using two or more optical systems in conjunction, some light rays
are lost because the pupils of the different systems do not match each
other. This effect is called vignetting. Generally, there are two
types of vignetting: spatial and angular vignetting.

\begin{figure}[h]
 \begin{center}
   \includegraphics[width=\linewidth]{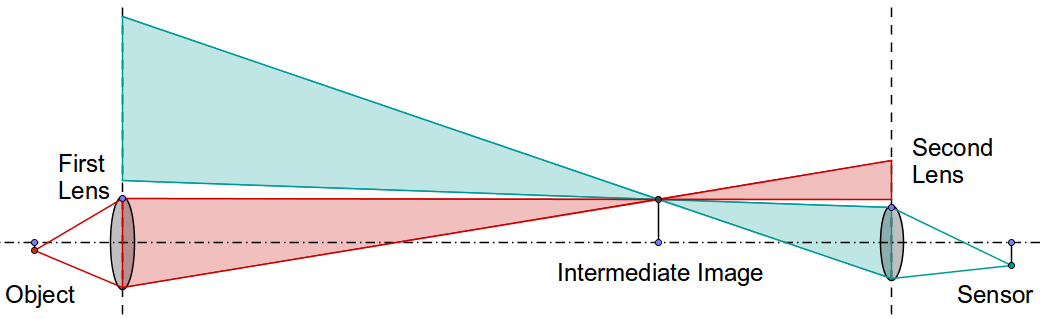}
   \includegraphics[width=\linewidth]{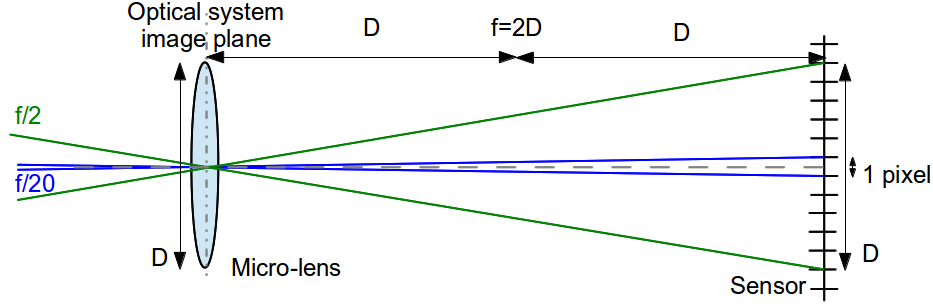}
 \end{center}
 \vspace{-3mm}
 \caption{{\em Top:} Spatial vignetting
 occurs when light rays from the object (in red) do not pass through
 the second lens. For non-vignetted imaging, light rays (in blue)
 converging to a point at the sensor edge should include all the red light
 rays emerging from the object. {\em Bottom:}
 Representation of the sampling of a cone of light rays by a $\fn=2$
 micro-lens. Green rays symbolize the angular cone that can be
 acquired by the micro-lens, while blue rays emerge from an
 $\fn\!=\!20$ optical system. As can be seen
 from the figure, angular vignetting prevents an effective sensor
 utilization. }
 \label{fig:vignetting}
\end{figure}

Spatial vignetting directly translates into a loss of field of view,
which may reduce the image size at the sensor (see Fig.~\ref{fig:vignetting}~(top)). Angular vignetting (see
Fig.~\ref{fig:vignetting}~(bottom)) occurs when the
cone of rays permitted through one of the systems is smaller than for
the other system, e.g. due to a stop positioned inside the
system. Angular vignetting is not an issue in a standard camera : it
only affects the exposure and is directly linked to the depth of field
of the camera. In a light-field camera, however, it is crucial to
minimize angular vignetting in order to prevent the loss of
directional light-field information.

We propose to measure the vignetting in terms of its adaptation to the
recording light-field sensor. An ideal optical system that is adapted to a
particular sensor would fully cover all its sensor elements. Since the
raw pixel resolution is divided into spatial and angular parts, the
sensor coverage $c_{sensor}$ can be approximately expressed as
%
\begin{equation}
c_{sensor} [\%] = c_{spatial} [\%] \times c_{angular} [\%],
\label{eq:coverage}
\end{equation}
%
%
where $c_{spatial}$ is the spatial coverage of the sensor in percent,
and $c_{angular}$ is the angular coverage of one micro-lens
sub-image, also in percent. In our experimental validation, we measure
the spatial coverage in the center view and the angular coverage in
the center lenslet. This choice is motivated by the simpler estimation
of the relevant coverage areas as compared to using the side
views/edges of the field.
The measure can be considered to be an approximation of the
upper bound of the system space-bandwidth product, i.e. the optical
information capacity of the system~\cite{Lohmann:96}.

\subsection{Unmodified Use of the Lytro's Main Optics}
\label{sec:lytro_opt}
The main optics of the Lytro camera, i.e. the optics without the
micro-lens array, is designed to avoid angular vignetting when imaging
onto the micro-lens array, i.e. the micro-lens array and the main optics have been designed with the same
f-number of $\fn=2$. We have observed that the
main optical system can be used in two different ways with a
microscope.  These two imaging regimes can be used differently in
designing an optical matching system.\\

\begin{figure}[h]
\centering
 \includegraphics[width=\columnwidth]{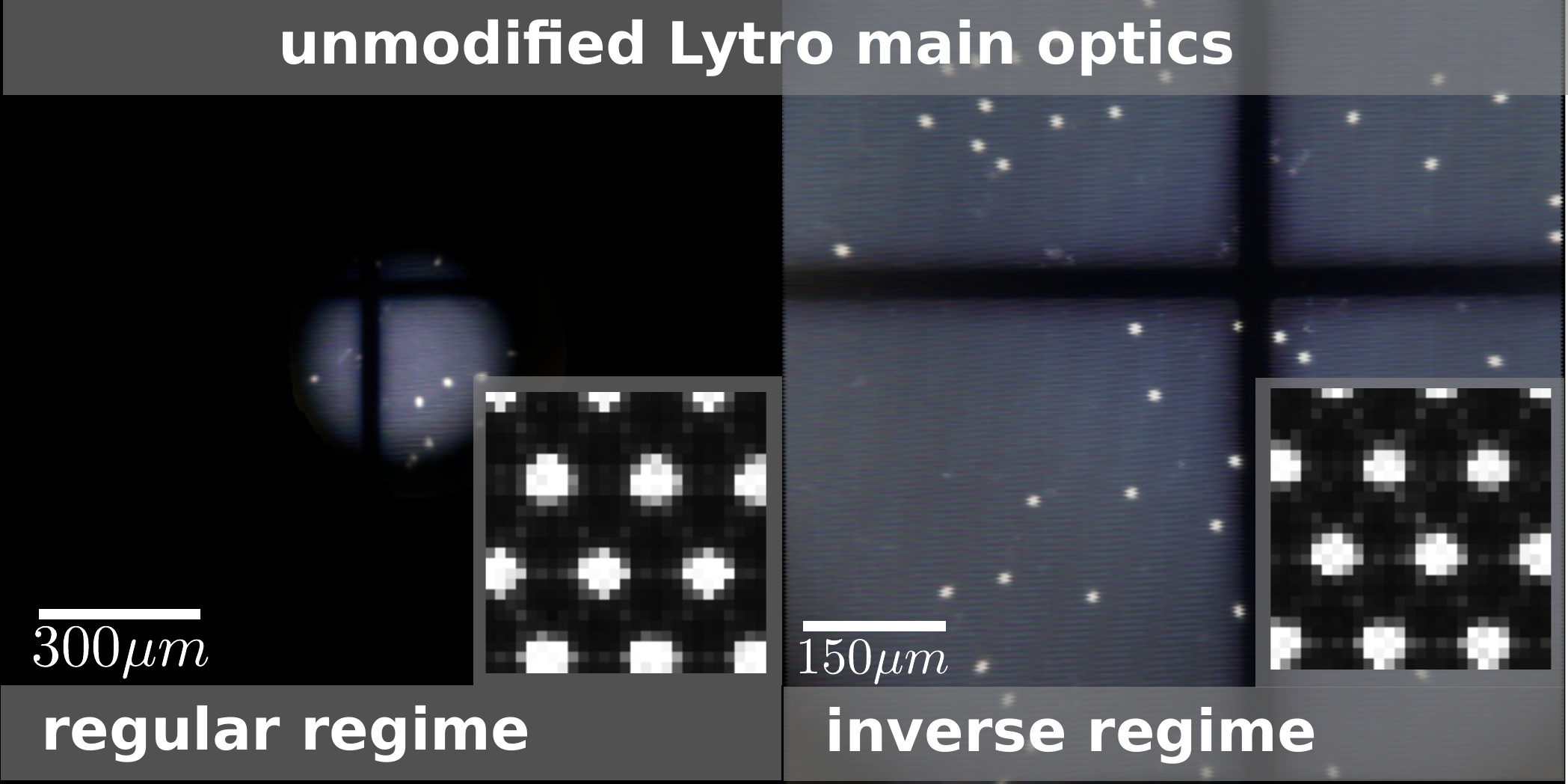}
 \vspace{-3mm}
 \caption{Direct imaging through a $10\times$
   microscope objective with an unmodified Lytro. The object is a blue LCD
   display. It consists of a square black grid that is separating the
   different pixels of $0.5\mm \times 0.5\mm$ size. We
   hypothesize the white dots to be bubbles inside the liquid crystal. The
   large images show the equivalent camera image computed from the
   light-field, while the small images on the bottom right show a
   close-up on the micro-lens images of the raw sensor data for
   different zoom-levels. Spatial and angular vignetting are easily
   observed in the equivalent camera and micro-lens images,
   respectively. The zoom level setting is $1\times$ (regular regime)
   on the left and $5\times$ (inverse regime) on the right. The
   spatial vignetting is strong in the regular regime (6\% of spatial
   coverage), while it is greatly improved in the inverse regime (100\%
   of spatial coverage). Angular coverage is similar in both
   cases ($c_{angular}\approx25\%$). }
\label{fig:vignetting_effect}
\end{figure}

\textbf{Regular regime:}
The Lytro camera can image a plane as close as the first surface of
its optics for a zoom level of $1\times$. This minimal focus distance
increases with the zoom level. In order to use the Lytro with the
microscope, it has to be positioned such that the near focus of the
camera is placed at the image plane of the microscope objective.
Since the image size $y'$ of the microscope objective is rather large (typically around $50\mm \times50mm$) whereas the Lytro's
entrance pupil is only $\approx 20 mm$ in
diameter, spatial vignetting incurs a loss of sensor
coverage of up to 94\% as shown in
Figure~\ref{fig:vignetting_effect}~(left).
The angular vignetting is stronger with only 16\% angular coverage.\\

\begin{figure}[h]
  \begin{center}
  \includegraphics[width=\columnwidth]{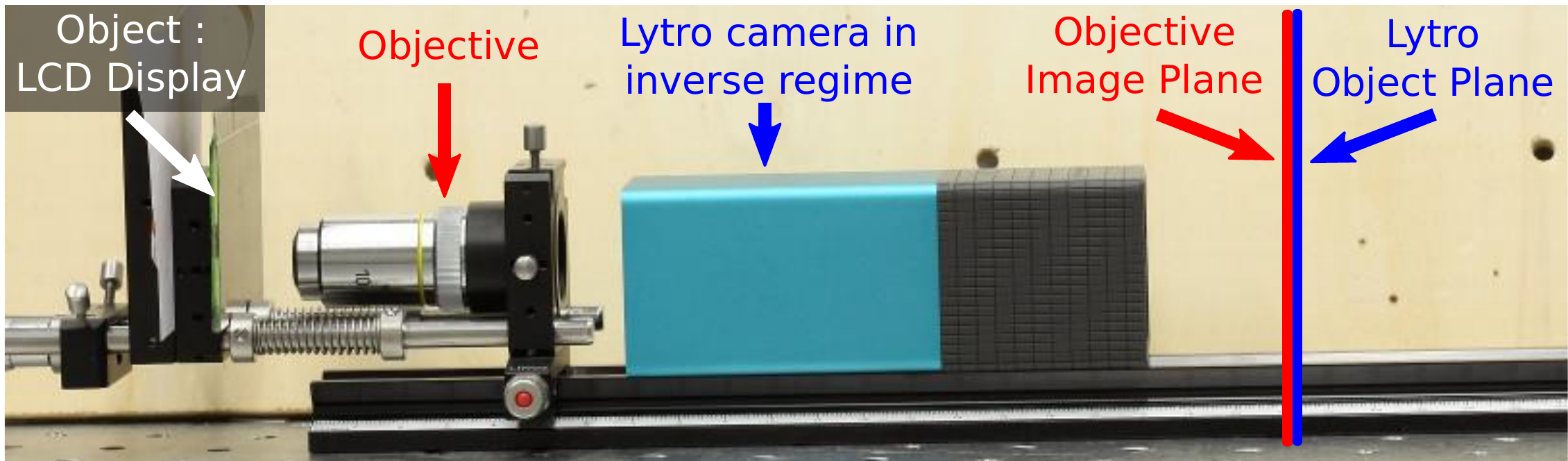}
  \includegraphics[width=\columnwidth]{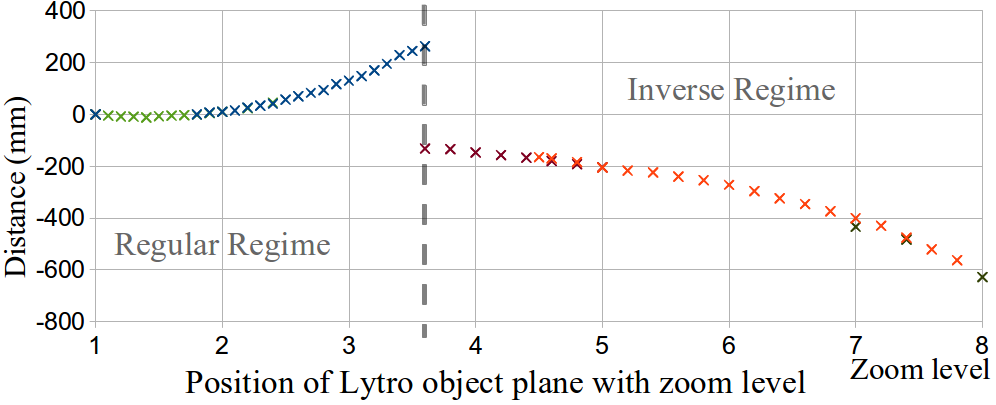}
 \end{center}
 \vspace{-3mm}
 \caption{\label{fig:lytro_micro}{\em Top:} Setup for the inverse regime configuration. {\em Bottom:} Distance of Lytro focus plane to its front lens with the variation of zoom. An abrupt change of position of the zoom lens group occurs for a zoom level of $3.7\times$, enabling to switch between the regular and inverse regime.}
\end{figure}

\textbf{Inverse regime:}
We discovered that, in a specific
configuration where the camera is set to focus to the closest possible
plane for a zoom level of~$1\times$, the camera can enter into a virtual
object regime. The camera is then able to image an object plane that is
located  {\em behind} its first lens, i.e. in the direction of the sensor (see Fig.~\ref{fig:lytro_micro}). This configuration enables the
positioning of the camera close to the microscope objective and
therefore reduces the spatial vignetting since a larger number of rays
can be captured by the lens surface (see
Fig.~\ref{fig:vignetting_effect}~(right). This mode of
operation inverts the image.
\paragraph{•} For both imaging regimes described above, the magnification $\magn_{Lytro}$ ($\approx$ 0.5) is not low enough for achieving a good
angular coverage. While the spatial vignetting problem can be
successfully addressed with the inverse regime, the angular
vignetting can only be dealt with by using a low magnification optical system. We therefore investigate the different options.

\section{The Lytro Microscope}
\label{sec:impladdoptics}

Our first option to achieve the matching of the f-numbers discussed
in Section \ref{sec:mismatch}, consists in designing an optical
demagnification system (placed between the microscope objective and 
the Lytro) that increases the angular extent of the light (see Eq.~\ref{equ:Abbe}).
This solution keeps desirable properties like the
large numerical aperture of the microscope and its fixed working
distance, while at the same time, the Lytro camera can remain
unmodified. The major task is to find a good trade-off between the
vignetting and the magnification of the resulting light-field
microscope.

\begin{figure}[h]
 \begin{center}
   \includegraphics[width=\linewidth]{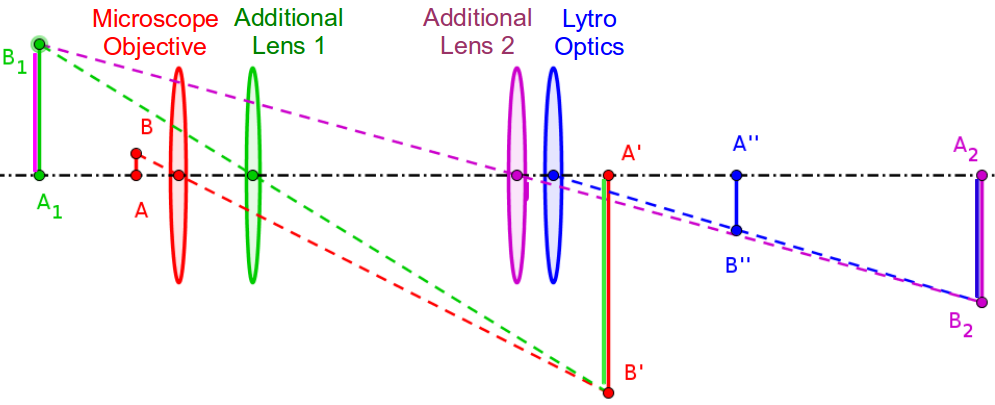}
 \end{center}
 \vspace{-3mm}
 \caption{\label{fig:ray-diagram2}Ray-diagram of the system using two additional lenses. 
 Each sub-system is indicated in a different color and has to be interpreted independently 
 from the other sub-systems. Their operation can be understood in sequence: the objective 
 images object $AB$ to $A'B'$, the first additional lens images $A'B'$ to $A_1B_1$ so that 
 the second lens images $A_1B_1$ to $A_2B_2$ behind the Lytro camera. Finally, the Lytro's main optics in the inverse regime
 images this virtual object $A_2B_2$ to its sensor in the plane $A''B''$.}
\end{figure}

Our best solution along this direction employs two lenses. This configuration serves two goals: 1) to decrease the magnification successively, simplifying the task of each individual lens, and 2) to move the image behind the Lytro camera so that it can
be used in the inverse regime which offers a better spatial coverage
$c_{spatial}$.
The ray-diagram in Fig.~\ref{fig:ray-diagram2} illustrates the two-lens
setting: an intermediate image that is slightly demagnified is created
in front of the microscope. Then, the second lens creates a further
demagnified image behind the Lytro camera. The Lytro camera is
operated in its inverse imaging regime in order to pick up this
virtual image.
We determine the positions and focal lengths of the additional lenses 
using the following equation derived from the thin lens formula:
\vspace{-2mm}
\begin{equation}
x= \fl \dfrac{\magn-1}{\magn}
\vspace{-2mm}
\label{equ:relation}
\end{equation}
The focal length $\fl$ and the position of the lens $x$ are chosen
according to the desired magnification $\magn$ (negative because the
image is inverted). In our implementation, the first additional lens has
a focal length of $50\mm$ and is put close to the microscope
objective. The second additional lens has a focal length of $85\mm$
and is put close to the Lytro.
The effect of using two lenses is that the individual focal lengths are
larger and the aberrations are reduced.
%
%

\section{Macro Lytro}
\label{sec:implrmoptics}
\label{sec:slr_lens}

Our second option is to use a single SLR camera lens in front of the Lytro to achieve unit magnification. ($\magn_{final}=\magn_{SLR}\magn_{Lytro}=1$). As for the microscope, and because the focal length of the SLR lens is large (50mm and 100mm in our experiments), we want to use the Lytro in the inverse regime to keep the spatial coverage as high as possible. In practice, the Lytro is set as close as possible to the SLR lens.
A variation of this setup is to remove the Lytro optics, and only use the SLR lens so that $\magn_{final}=\magn_{SLR}=1$.
These designs have only one or two optical
components and relieve the hurdle of undoing the work of the microscope
objective with many lenses. However, the SLR lens is not specifically
designed for the magnification of close objects and its aperture is
not meant to be maintained at a constant value for the micro-lens
array. The relations used to establish Eq.~\ref{equ:microscope_fnumber} are not valid in this macro-configuration. Instead, the relevant quantity is the working f-number $\fn_w$~\cite{born1999principles}
which is the f-number modified by the magnification $\magn$:
\vspace{-2mm}
\begin{equation}
  \fn_w= \dfrac{1}{2\na_i} = (1-\magn)\fn
\vspace{-2mm}
\label{equ:working_fnumber}  
\end{equation}
where $\na_i$ is the image-side numerical aperture $\na_i= \ri'
sin\alpha'$.  The minimal value of working f-number that can be
achieved with a camera lens is close to $\fn_w=2$. It is reached for a
limit f-number of $\fn=1$ and a magnification of $\magn=-1$. This
condition would be optimal for the Lytro micro-lens array. However,
commercial lenses usually have a limit f-number between 1.4 and 3.5
increasing the working f-number to between 2.8 and 7.0.

Compared to the light-field microscope, on the one hand, this setup is
more versatile. The magnification can be set to the desired value by
simply moving the object and the micro-lens array. It does not require
the difficult alignment of several optics. 
On the other hand, this setup does not benefit from the
structure of the microscope that already includes lighting and moving
the sample through micrometer stages in three dimensions. 
In addition, the working
distance is not fixed which changes the magnification as well as the
object-side numerical aperture when moving the sample.  However, the
strong point of this design is its accessibility. Building a
light-field macrography setup is done quickly and without the need
for a deep understanding of the operating principles of a microscope.

\section{Results}
\label{sec:results}

\begin{figure}[h]
 \begin{center}
   \includegraphics[width=\linewidth]{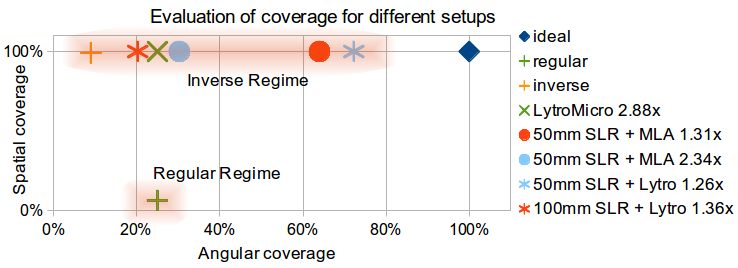}
 \end{center}
 \vspace{-3mm}
 \caption{Combined results of the experiments from
   Sect.~\ref{sec:impladdoptics} and Sect.~\ref{sec:implrmoptics}.}
\label{fig:vignetting_diagram}
\end{figure}

Before showing results, we describe and compare the different implementations. The
different sensor coverages as well as their spatio-angular
coverage values can be found in Fig.~\ref{fig:vignetting_diagram}. It is
clearly visible that directly using the Lytro camera in its regular
imaging regime is unsuitable for microscopic light-field imaging. The inverse
imaging regime improves on the spatial coverage, but the angular
coverage is limited. The best combination is achieved with a 50mm SLR lens
in front of the Lytro which yields the best overall sensor coverage $c_{sensor}$.

\subsection{Resolution Test Chart}
\label{sec:comparison}
\begin{figure}[h]
 \begin{center}
   \includegraphics[width=\linewidth]{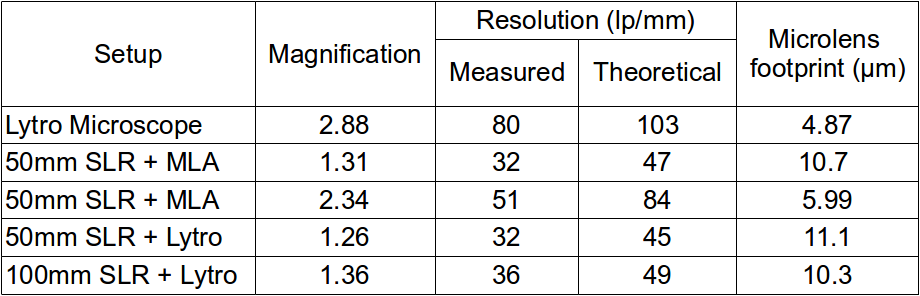}
 \end{center}
 \vspace{-3mm}
 \caption{\label{fig:tabres}This table summarizes the results from the experiments
 of Sect.~\ref{sec:impladdoptics} and Sect.~\ref{sec:slr_lens}. Theoretical
 resolution is derived from the micro-lens pitch. The micro-lens footprint is the size of a micro-lens in object space.}
\end{figure}

In order to compare the resolution of the two different techniques, we
use a 1951 USAF resolution test chart. The results of the experiments are summarized in Fig.~\ref{fig:tabres}.

\begin{figure*}[t]
 \begin{center}
   \includegraphics[width=\linewidth]{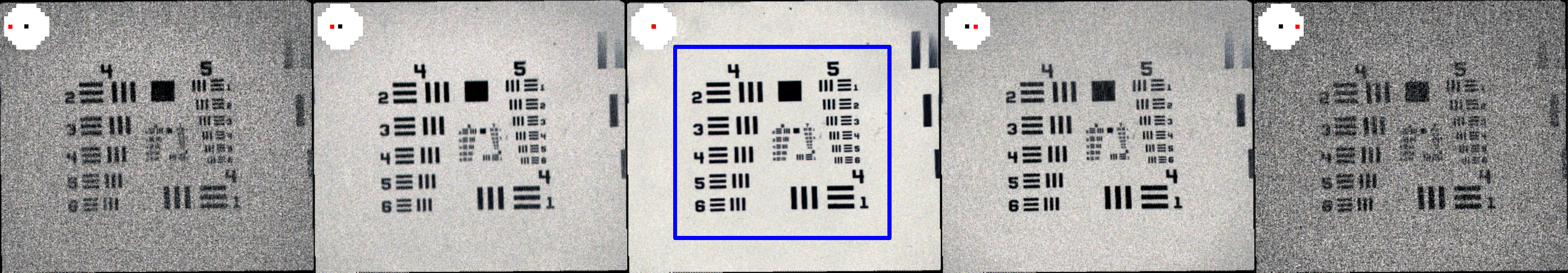} 
   \includegraphics[width=\linewidth]{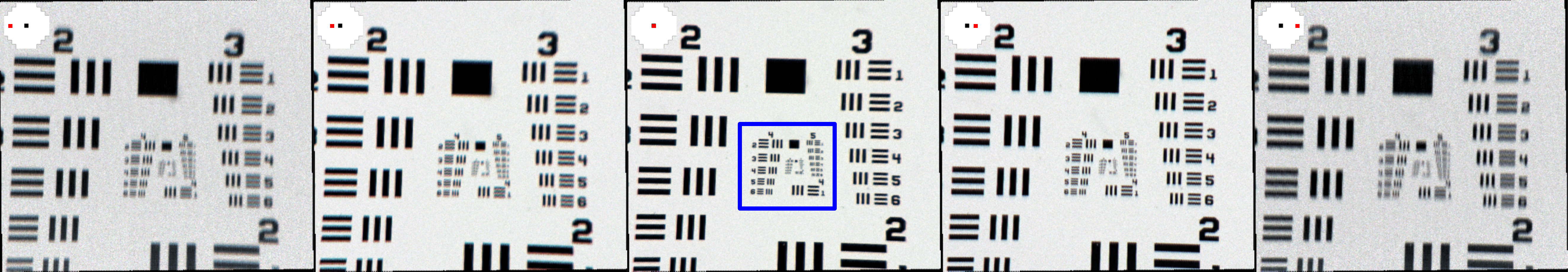} 
 \end{center}
 \vspace{-3mm}
 \caption{ \label{fig:results_target} Set of different viewpoints of
the resolution target with the center view in the middle (the red dot in the top left inset indicates the position of the view). {\em Top:}
Images taken with the "Lytro Microscope" (Sect.~\ref{sec:impladdoptics}). The
magnification is 2.88. {\em Bottom:} Images taken with the "50\mm SLR + Lytro" (Sect.~\ref{sec:slr_lens}). A red-green color shift due to
strong chromatic aberrations can be observed in the side
views. Note that the top row has a higher resolution: it shows
the pattern that is visible in the center of the bottom row (level 4
and 5). The contrast of the images of the same row have been set to a similar level for comparison.}
\vspace{-3mm}
\end{figure*}

The first option (Sect.~\ref{sec:impladdoptics}) "Lytro Microscope" was implemented using
a Canon $50\mm$ SLR lens and a Nikon $85\mm$ SLR lens as additional
lenses. Those lenses were put on top of a Leitz Ergolux microscope using an objective of magnification $10\times$ with an
object-side numerical aperture $\na_o=0.2$. This microscope has a lens tube
with a magnification of $0.8\times$ so the f-number $\fn=20$.
The images have been taken with a magnification of 2.88, i.e. a micro-lens covers $4.87\mum$ in
object space (see Fig.~\ref{fig:results_target}~(top)).
The spatial coverage is above 99\% but due to the large magnification the angular
coverage is low (between $9\%$ and $25\%$). The resolution is between
80 and 90 line pairs per $\mm$.

The resolution indicated above is computed for the center
view. It decreases with further distance from the center. A loss
of image quality due to aberrations can be observed. They are
introduced because the observed area is larger than usual for the
microscope objective. Microscope objectives are typically designed so
that only a reduced inner portion of the full field is very well
corrected. In addition, our matching lenses introduce further
aberrations.
Since the angular vignetting is strong, the contrast of viewpoints far
from the center is low. It should be noted that even viewpoints inside
the vignetted area can be computed, albeit at a poor
signal-to-noise ratio (see Fig.~\ref{fig:results_target}).

\begin{figure}[h]
 \begin{center}
   \includegraphics[width=\linewidth]{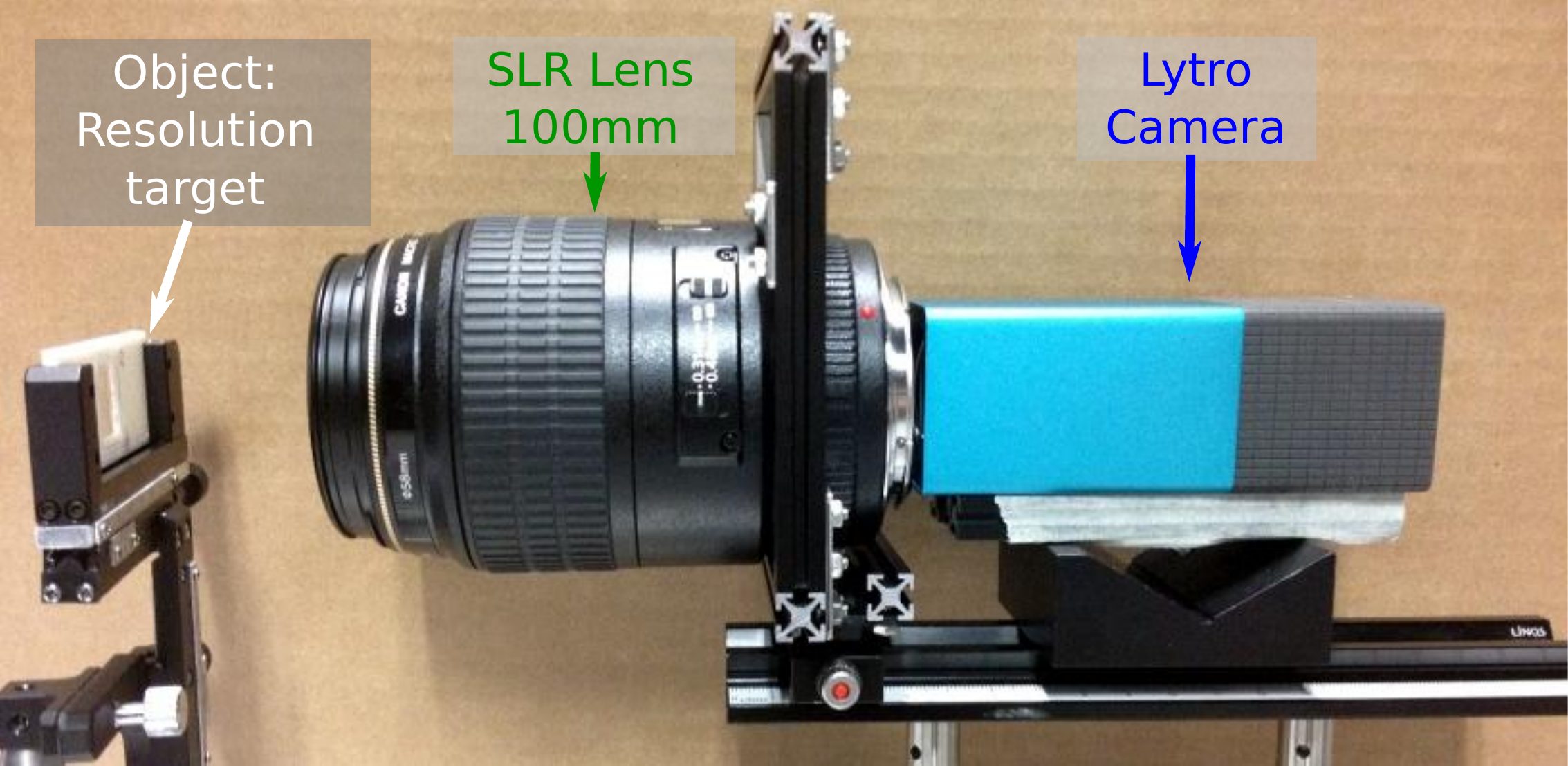}
 \end{center}
 \vspace{-3mm}
 \caption{Setup from the SLR + Lytro experiment using a 100mm SLR lens (100mm SLR + Lytro).}
 \label{fig:1to1setup}
\end{figure}

The second option (Sect.~\ref{sec:slr_lens}) was implemented in three
ways: two times with the Lytro placed behind two different lenses, a
$50\mm$ and a $100\mm$ Canon SLR lens (referred as SLR + Lytro), and
with the Lytro micro-lens array without the Lytro main optics, see
Fig.~\ref{fig:mla_closeup2} (right), placed behind the $50\mm$ lens
(referred as SLR + MLA) (see Fig.~\ref{fig:1to1setup}).
Magnifications from 1.26 to 2.34 were achieved. The spatial coverage
is always 100\% and angular coverage is good (up to $70\%$).
In this case, chromatic aberrations are present which degrades the
image. The aberrations are reduced in the SLR + Lytro case as compared
to SLR + MLA, since the magnification of the SLR lens is lower in this
setting. It is most noticable in the side views since, for these
views, imaging is performed through the outer pupil regions of the SLR
lens. We suspect that the chromatic aberration is introduced by the
SLR lens because it is not intended for macro-imaging. Using a
dedicated macro-lens instead would likely remove this effect.


\subsection{Microscopic Sample}
\label{sec:applications}
The most direct application of the light-field microscope is the study of
microscopic samples. The low magnification and the large field of view allow us
to see in detail an object area that is between $1.5\mm\times 1.5\mm$ and
$3.5\mm\times 3.5\mm$ with a magnification of $2.88\times$ and $1.3\times$
respectively. Cell tissues or rough surfaces of different materials have a
structure close to the millimeter so high magnification is not always necessary
to analyze them.

We illustrate this technique in
Fig.~\ref{fig:samples_results}~(right). Several images of microscopic
specimen were taken with the same settings as in the previous
section. The magnification is $2.88\times$ and we can clearly see the
structure of different kind of surfaces that are invisible to the
naked eye.

\begin{figure}[h]
 \begin{center}
   \includegraphics[width=\linewidth]{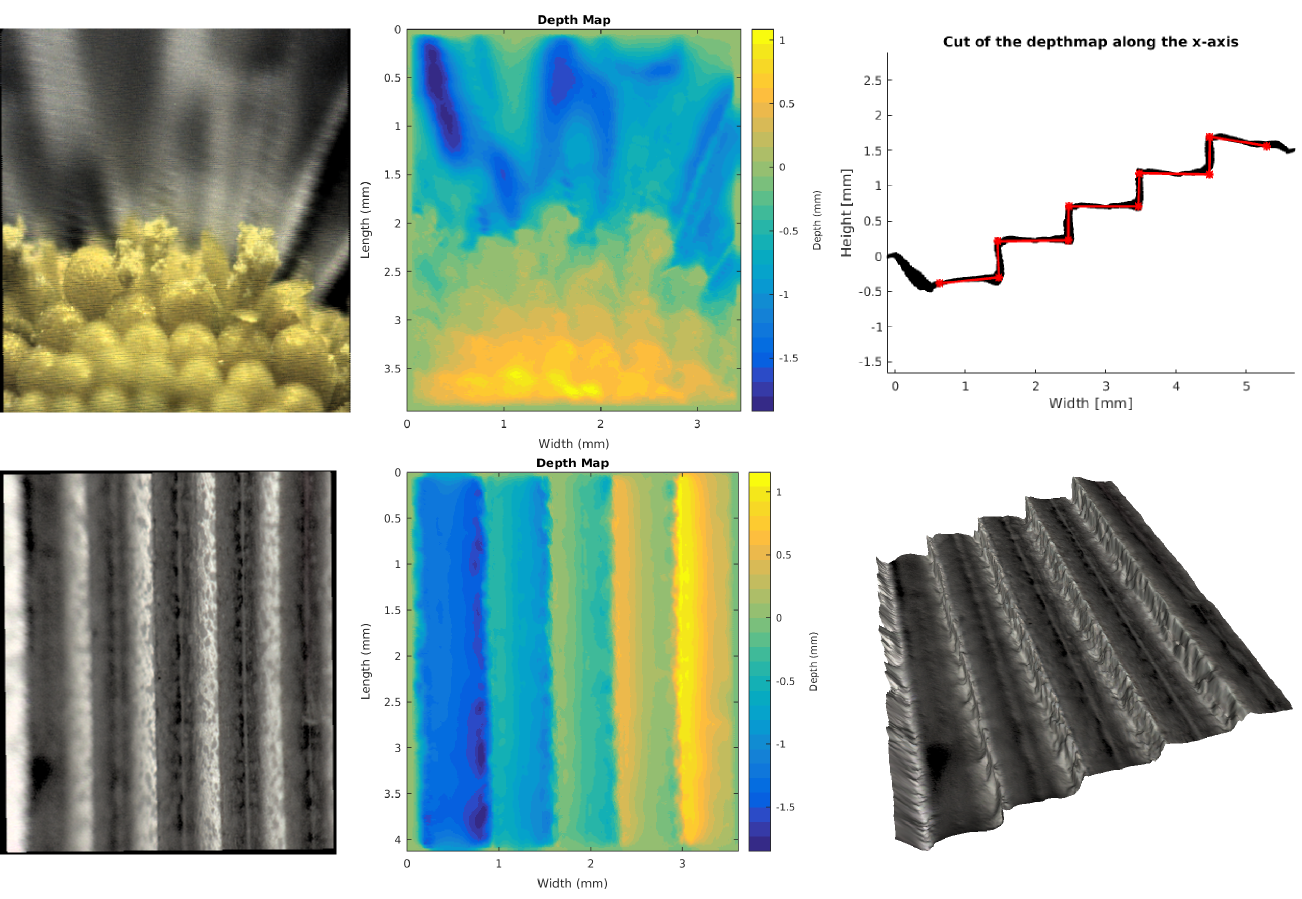}
 \end{center}
	\vspace{-4mm}
 \caption{ \label{fig:daisy_depthmap}Sub-view (top left) and computed depth map
 (top middle) of a daisy flower. Sub-view (bottom left) and computed depth map
 (bottom middle) of the aluminium staircase. The bottom right picture is a 3D
 visualization of the depth map after the calibrating and the top right picture
 is a projection of a region in the middle of the calibrated depth map onto the
 xz plane showing the plane fits in red.}
	\vspace{-5mm}
\end{figure}

The light-field data allows for the reconstruction of the depth of the
sample when the number of views is sufficient. We took a picture of a
ground truth aluminium staircase with stairs of $1.00\mm$ width and
$0.50\mm$ height with an accuracy of $\pm5\mum$ with the 50\mm SLR + Lytro
setup. We obtained the depth map in Fig.~\ref{fig:daisy_depthmap}
using a modified variational multi-scale optical flow algorithm for
light-field depth estimation~\cite{Manakov:13}. Although only a small
slice of the staircase is in focus (the depth of field is $1\mm$), the
depth can be computed outside of this area.
Essentially, out-of-focus regions are naturally considered as a different
scale by the algorithm, so, the estimation of the depth of the closest and furthest steps is correct.
This behavior nicely interacts with the scene properties since the parallax
is larger in out-of-focus regions. The optical system can be seen as
supporting the part of the algorithm that is handling large
displacements.
Since the detailed properties of the Lytro main optics are unknown,
it is necessary to adjust the scale of the computed depth. We use the aluminium
staircase as reference.
We find the affine transformation between the depth map and the staircase
model by fitting planes to match the stairs. After the transformation,
we measure an RMS error of
$75\mum$ for vertical planes and $17\mum$ for horizontal planes.
The difference is due to the different slopes of the horizontal and
vertical steps with relation to the camera view direction.
The magnification is equal to 1.32.
We also applied this depth reconstruction to a daisy head (see Fig.~\ref{fig:daisy_depthmap}).
In practice, the depth inside a cube of about
$3.5\times3.5\times3.5mm^{3}$ can be estimated, which is rather large
for a microscopic setting.


%

\section{Discussion and Conclusion}
\label{sec:conclusions}

We have developed and tested several adaptations of the Lytro consumer
light-field camera to enable an entry-level experimentation with light-field 
microscopy. While the fixed f-number of the Lytro's micro-lens
array prevents its direct use with a standard microscope (regular
regime), it is possible to trade the overall system magnification for
light-field features and to avoid spatial vignetting with the inverse
imaging regime.  Lytro microscopy is therefore an option for
low-magnification work as is common in industrial settings, or for
investigations into the meso- and large-scale micro-structure of
materials.  Even though an optical magnification between $1$ and $3$, as
achieved in this work, appears to be low, the small size of the
micro-lenses still yields a decent optical resolution of up to $6.25
\mum$ in object space which already shows interesting optical structures that are imperceivable by the naked eye.

For the future, we would like to investigate image-based denoising
schemes for the vignetted side-views, as well as algorithmic
developments for structure recovery. In terms of applications, the
imaging of micro-and meso-BRDFs and their relation to macroscopic BRDF
models appears to be an interesting development. 

\todo{Remove all TODOs  and ivos}

\section*{Acknowledgements}
We would also like to thank Patrick Reuter for his helpful comments.
This work was supported by the German Research Foundation (DFG)
through Emmy-Noether fellowship IH 114/1-1 as well as the ANR ISAR project of the French Agence Nationale de la Recherche.
{\small
\bibliographystyle{ieee}
\bibliography{references}

\begin{thebibliography}{10}\itemsep=-1pt

\bibitem{born1999principles}
M.~Born and E.~Wolf.
\newblock {\em Principles of optics}.
\newblock Pergamon Press, 1980.

\bibitem{Broxton:13}
M.~Broxton, L.~Grosenick, S.~Yang, N.~Cohen, A.~Andalman, K.~Deisseroth, and
  M.~Levoy.
\newblock {Wave Optics Theory and 3-D Deconvolution for the Light Field
  Microscope}.
\newblock {\em {Optics Express}}, 21(21):25418--25439, 2013.

\bibitem{cohen2014enhancing}
N.~Cohen, S.~Yang, A.~Andalman, M.~Broxton, L.~Grosenick, K.~Deisseroth,
  M.~Horowitz, and M.~Levoy.
\newblock Enhancing the performance of the light field microscope using
  wavefront coding.
\newblock {\em Optics express}, 22(20):24817--24839, 2014.

\bibitem{georgiev2006light}
T.~Georgiev and C.~Intwala.
\newblock Light field camera design for integral view photography.
\newblock Technical report, Adobe System, Inc, 2006.

\bibitem{ihrke2008fast}
I.~Ihrke, T.~Stich, H.~Gottschlich, M.~Magnor, and H.-P. Seidel.
\newblock Fast incident light field acquisition and rendering.
\newblock {\em Journal of WSCG (WSCG'08)}, 16(1-3):25--32, 2008.

\bibitem{miloushnet}
J.~Ku\u{c}era.
\newblock Lytro meltdown.
\newblock \url{http://optics.miloush.net/lytro/Default.aspx}, 2014.

\bibitem{lanman2006spherical}
D.~Lanman, D.~Crispell, M.~Wachs, and G.~Taubin.
\newblock Spherical catadioptric arrays: Construction, multi-view geometry, and
  calibration.
\newblock In {\em 3D Data Processing, Visualization, and Transmission, Third
  International Symposium on}, pages 81--88. IEEE, 2006.

\bibitem{Lee:12}
J.-J. Lee, D.~Shin, B.-G. Lee, and H.~Yoo.
\newblock {3D Optical Microscopy Method based on Synthetic Aperture Integral
  Imaging}.
\newblock {\em {3D Research}}, 3(4):1--6, 2012.

\bibitem{levoy2006light}
M.~Levoy, R.~Ng, A.~Adams, M.~Footer, and M.~Horowitz.
\newblock Light field microscopy.
\newblock In {\em ACM Transactions on Graphics (TOG)}, volume~25, pages
  924--934. ACM, 2006.

\bibitem{Levoy:09}
M.~Levoy, Z.~Zhang, and I.~McDowall.
\newblock {Recording and Controlling the 4D Light Field in a Microscope using
  Microlens Arrays}.
\newblock {\em {Journal of Microscopy}}, 235:144--162, 2009.

\bibitem{liang2008programmable}
C.-K. Liang, T.-H. Lin, B.-Y. Wong, C.~Liu, and H.~H. Chen.
\newblock Programmable aperture photography: Multiplexed light field
  acquisition.
\newblock {\em ACM Trans. on Graphics}, 27(3):1--10, 2008.

\bibitem{lippman1908epreuves}
G.~Lippman.
\newblock {{\'E}preuves r{\'e}versibles photographies int{\'e}grales}.
\newblock {\em CR Acad. Sci}, 146:446--451, 1908.

\bibitem{Lohmann:96}
A.~W. Lohmann, R.~G. Dorsch, D.~Mendlovic, Z.~Zalevsky, and C.~Ferreira.
\newblock {Space-Bandwidth Product of Optical Signals and Systems}.
\newblock {\em {Journal of the Optical Society of America A}}, 13(3):470--473,
  1996.

\bibitem{Lu:13}
C.-H. Lu, S.~Muenzel, and J.~Fleischer.
\newblock {High-resolution Light-Field Microscopy}.
\newblock In {\em Computational Optical Sensing and Imaging}, pages CTh3B--2.
  Optical Society of America, 2013.

\bibitem{Manakov:13}
A.~Manakov, J.~Restrepo, O.~Klehm, R.~Heged{\"u}s, H.-P. Seidel, E.~Eisemann,
  and I.~Ihrke.
\newblock {A Reconfigurable Camera Add-On for High Dynamic Range,
  Multispectral, Polarization, and Light-Field Imaging}.
\newblock {\em ACM Trans. on Graphics (SIGGRAPH'13)}, 32(4):article 47, 2013.

\bibitem{Murphy:12}
D.~B. Murphy and M.~W. Davidson.
\newblock {\em {Fundamentals of Light Microscopy and Electronic Imaging}}.
\newblock John Wiley \& Sons, 2012.

\bibitem{ng2005light}
R.~Ng, M.~Levoy, M.~Br{\'e}dif, G.~Duval, M.~Horowitz, and P.~Hanrahan.
\newblock Light field photography with a hand-held plenoptic camera.
\newblock {\em Computer Science Technical Report CSTR}, 2(11), 2005.

\bibitem{Oldenbourg:08}
R.~Oldenbourg.
\newblock {Polarized Light Field Microscopy: An Analytical Method using a
  Microlens Array to Simultaneously Capture both Conoscopic and Orthoscopic
  Views of Birefringent Objects}.
\newblock {\em {Journal of Microscopy}}, 231(3):419--432, 2008.

\bibitem{taguchi2010axial}
Y.~Taguchi, A.~Agrawal, S.~Ramalingam, and A.~Veeraraghavan.
\newblock Axial light field for curved mirrors: Reflect your perspective, widen
  your view.
\newblock In {\em Computer Vision and Pattern Recognition (CVPR), 2010 IEEE
  Conference on}, pages 499--506. IEEE, 2010.

\bibitem{Thomas:13}
M.~Thomas, I.~Montilla, J.~Marichal-Hernandez, J.~Fernandez-Valdivia,
  J.~Trujillo-Sevilla, and J.~Rodriguez-Ramos.
\newblock {Depth Map Extraction from Light Field Microscopes}.
\newblock In {\em 12th Workshop on Information Optics (WIO)}, pages 1--3. IEEE,
  2013.

\bibitem{Veeraraghavan:08}
A.~Veeraraghavan, R.~Raskar, A.~Agrawal, R.~Chellappa, A.~Mohan, and
  J.~Tumblin.
\newblock {N}on-{R}efractive {M}odulators for {E}ncoding and {C}apturing
  {S}cene {A}ppearance and {D}epth.
\newblock In {\em IEEE Conference on Computer Vision and Pattern Recognition
  (CVPR)}, pages 1--8, 2008.

\bibitem{Veeraraghavan:07}
A.~Veeraraghavan, R.~Raskar, A.~Agrawal, A.~Mohan, and J.~Tumblin.
\newblock {Dappled Photography: Mask Enhanced Cameras For Heterodyned Light
  Fields and Coded Aperture Refocussing}.
\newblock {\em ACM Trans. on Graphics (TOG)}, 26:69, 2007.

\bibitem{wilburn2005high}
B.~Wilburn, N.~Joshi, V.~Vaish, E.-V. Talvala, E.~Antunez, A.~Barth, A.~Adams,
  M.~Horowitz, and M.~Levoy.
\newblock High performance imaging using large camera arrays.
\newblock In {\em ACM Trans. on Graphics (TOG)}, volume~24, pages 765--776.
  ACM, 2005.

\bibitem{Wilt:09}
B.~A. Wilt, L.~D. Burns, E.~T.~W. Ho, K.~K. Ghosh, E.~A. Mukamel, and M.~J.
  Schnitzer.
\newblock {Advances in Light Microscopy for Neuroscience}.
\newblock {\em {Annual Review of Neuroscience}}, 32:435, 2009.

\bibitem{yang2002real}
J.~C. Yang, M.~Everett, C.~Buehler, and L.~McMillan.
\newblock A real-time distributed light field camera.
\newblock In {\em Proceedings of the 13th Eurographics workshop on Rendering},
  pages 77--86. Eurographics Association, 2002.

\end{thebibliography}
}

\end{document}